\documentclass[]{spie}  

\usepackage{array}
\usepackage{wrapfig}
\usepackage{multirow}
\usepackage{tabularx}
\usepackage{cite}
\usepackage{multicol}
\usepackage{balance}
\usepackage[colorlinks=true, allcolors=blue]{hyperref}
\usepackage[pdftex]{graphicx}
\usepackage{epstopdf}
\usepackage{subfigure}

\usepackage{pgfplots,multicol}
\usepackage{amsmath,stmaryrd,pict2e,picture}
\usepackage{amsmath,amssymb}
\usepackage{mathtools}
\usepackage{pifont}

\newlength{\rowidth}
\AtBeginDocument{\setlength{\rowidth}{3em}}
\usepackage{mathrsfs}
\usepackage{algorithm, algorithmic}

\def\footnoterule{\relax%
	\kern-5pt
	\hbox to \columnwidth{\hfill\vrule width 0.75\columnwidth height 0.4pt\hfill}
	\kern4.6pt}
\makeatother
\title{Multi Antenna Radar System for American Sign Language (ASL) Recognition Using Deep Learning}

\author{Gavin MacLaughlin}
\author{Jack Malcolm}
\author{Syed~A.~Hamza}

\affil{School of Engineering, Widener University, Chester, PA 19013, USA}

\authorinfo{E-mails: 	 \{gmaclaughlin, jfmalcolm, shamza\}@widener.edu
}
\begin{document}

\maketitle
\begin{abstract}
This paper investigates RF-based system for automatic American Sign Language (ASL) recognition.   We consider radar for ASL by joint spatio-temporal preprocessing of radar returns using time frequency (TF) analysis and high-resolution receive beamforming. The additional degrees of freedom offered by joint temporal and spatial processing using a multiple antenna sensor can help to recognize ASL conversation between two or more individuals. This is performed by applying beamforming to collect spatial images in an attempt to resolve individuals  communicating at the same time through hand and arm movements. The spatio-temporal images are fused and classified by a convolutional neural network (CNN) which is capable of discerning signs performed by different individuals even when the beamformer is unable to separate the respective  signs completely.  The focus group comprises  individuals with varying expertise with sign language, and real time measurements at 77 GHz frequency are performed using Texas Instruments (TI) cascade radar. 

\end{abstract}

\section{Introduction}

Radar systems present an attractive solution for human motion recognition (HMR) finding important applications in a large variety of scenarios ranging from gesture recognition for smart homes, detecting events of interest for automatic  surveillance, daily activities and behavioral analysis, Gait abnormality recognitions, health monitoring in  care facilities and  rehabilitation services   to enable independent living for elderly \cite{6126543, Amin2017RadarFI, 8610109, WANG2018118, 8613848, 9069251, 8746862}.

The main thrust of research for ASL recognition thus far has been on wearable or camera-based systems.  Camera based systems are prone to privacy concerns and are affected by lighting conditions whereas wearable sensor gloves are inconvenient to use for many practical purposes. Radar is a proven technology for  target detection, localization and tracking. It has been successfully used for hand and arm gesture recognitions \cite{9123903, 9114779, 8610109}. ASL involves both right and left hands and arms and as such are more intricate than individual hand or arm movements. This intricacy translates to MicroDoppler signatures with complex structures. It has been recently shown that radar can faithfully recognize sign language and provides privacy preserving/non-contact ASL recognition to improve the quality of life for a vulnerable group of individuals and will help the deaf and hard-of-hearing. High resolution radars are getting attention recently because of their added capability of   resolving closely spaced targets. In this case, radar  classification could be performed after localizing the target in range, Doppler and/or  angle.   We investigate radar system to concurrently discern sign language of more than one person and would  allow to  streamline the communications between the multiple  hard of hearing persons and voice activated devices. 

In this paper, we consider the ASL classification of  multiple persons by attempting to localize the signs to specific azimuth directions. The proposed beamforming approach can reduce the system cost and alleviate the need of using multiple radars as proposed in \cite{8010417, 8835796, 8835618,  article123, 10.1117/12.2588700}. Azimuth filtering is achieved by  applying  beamforming to the receiver array. Specifically, we consider the task of classifying two persons performing breathe, drink and come signs at different azimuth angles. We aim to correctly  pair the performed sign to  the corresponding azimuth angle. This is achieved by jointly processing  the received signals in the spatial and temporal domains. For a given scenario,  three different spectrograms are generated namely,  the combined spectrogram and the two other spectrograms obtained  through beamforming in the  two different azimuth directions. In this case, the received data is filtered using beamforming with two separate sets of beamformer weights. Subsequently, classification is performed by jointly processing all the three time frequency signatures.   The proposed scheme works adequately  when the two individuals are not completely separable in azimuth. This could either be due to the close proximity of the two persons in the azimuth or due to leakage associated with high sidelobes of the beamformer.   

\begin{figure*}[!t]
	\centering
	\includegraphics[height=2.5in, width=6.77in]{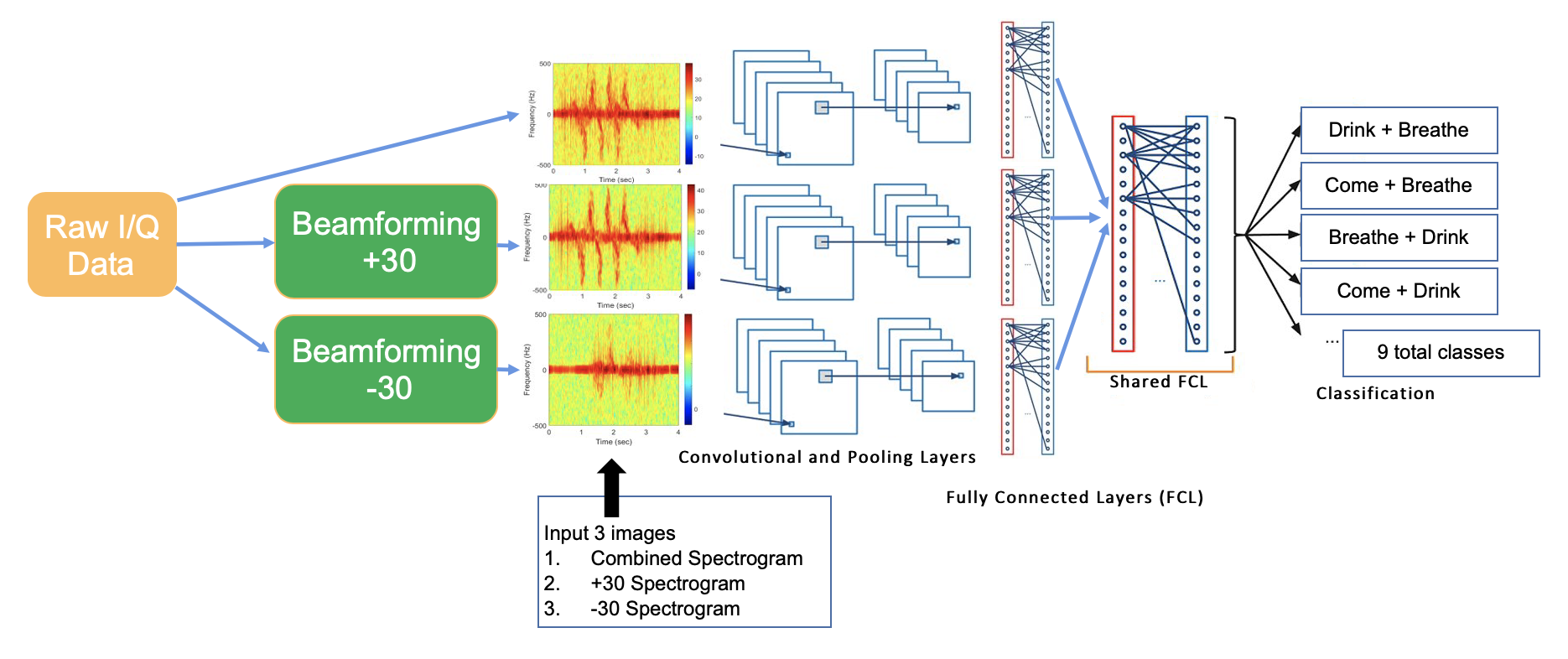}
	\caption{Processing sequence}
	\label{Fig.1}
\end{figure*} 
The rest of the paper is organized as follows:  In the next section, we  describe the  data set, beamforming and time frequency representation domain. The  beamformer image fusion  and convolutional neural network (CNN) for multi-person monitoring is discussed in Section \ref{Feature extraction and classification}. Experimental results are shown in Section \ref{Experimental Results}, while the conclusion is given in Section \ref{Conclusion}.
\section{Radar return signal analysis}
\label{Problem Formulation}
The complex valued raw  data matrix $\mathbf{s}(n,m) \in C^{N*M}$  of the frequency-modulated continuous
wave (FMCW) radar  is obtained through spatially processing the radar returns    by an  $M$ element uniformly spaced antenna array. The data is collected over  $N$ temporal sampling instances. The receiver array vector $\mathbf{s}(m) \in C^{M}$ at  time instant $n$ corresponds to the $n_{th}$ row of $\mathbf{s}(n,m)$ and is given by,
\begin {equation} \label{a}
\mathbf{s}(m)=    \sum_{l=1}^{L} \alpha _l \mathbf{a}( \theta_l)  + \mathbf{v}(m),
\end {equation}
where, $\mathbf{a} ({\theta_l})$  $\in \mathbb{C}^{M}$ is  the  steering vector   corresponding to the azimuth direction $\theta_l$ of the scatterer, and is defined  as follows,  
\vspace{+2mm}
\begin {equation}  \label{b}
\mathbf{a} ({\theta_l})=[1 \,  \,  \, e^{j (2 \pi / \lambda) d cos(\theta_l)  } \,  . \,   . \,  . \, e^{j (2 \pi / \lambda) d (M-1) cos(\theta_l)  }]^T.
\end {equation}
Here, $d$ is the inter-element spacing  and $\alpha_l$ $\in \mathbb{C}$  is the complex amplitude of the radar return. The additive Gaussian noise $\mathbf{v}(m)$ $\in \mathbb{C}^M$ has   variance   $\sigma_v^2$.
The elements of the received data vector $\mathbf{s}(m)$   are   combined linearly by the $M$-sensor  beamformer that strives to spatially filter the reflections from all other directions except  the signal in the direction of beamformer look angle $\theta_k$. The spatially filtered signal vector $\mathbf{x}({\theta_k})$ $\in \mathbb{C}^N$ after beamforming is given by, 
\begin {equation}  \label{c}
\mathbf{x}({\theta_k}) = \mathbf{s}(n,m) \mathbf{w}^H({\theta_k}),
\end {equation}
where $\mathbf{w}({\theta_k})=\mathbf{a}^H ({\theta_k})$ are the complex beamformer weights pointing towards $\theta_k$.

The spatially filtered signal vector $\mathbf{x}({\theta_k})$  is reshaped into a two-dimensional matrix, $\mathbf{x}_{\theta_k}(p, q)$. This is achieved  by segmenting the $N$ dimensional vector $\mathbf{x}({\theta_k})$, such that, the $P$ samples collected within a pulse repetition interval (PRI) are stacked into a $P$ dimensional column. There are   $Q$ such columns  within $\mathbf{x}_\theta(p, q)$ where  $Q=N/P$  is the number of PRIs processed within the observation time $N$.  The range-map, $\mathbf{r}_{\theta_k}(p,q)$  is obtained by applying the column-wise Discrete Fourier Transform (DFT) operation which is given by,
\begin {equation}  \label{d}
\mathbf{r}_{\theta_k}(l,q) =  \sum_{p=0}^{P-1} \mathbf{x}_{\theta_k}(p, q)e^{-j (2 \pi l p/ N)}
\end {equation}
We observe the data in the TF domain after localizing the motion in azimuth and range bins of interest. The spectrogram is used as the TF signal representation, showing the variation of the signal power as a function of  time $n$ and frequency $k$. The spectrogram of a periodic version of
a discrete signal $\mathbf{v}_{\theta_k}(n)$, is given by \cite{30749, article4573,  9101078}, 
\begin {equation}  \label{e}
\mathbf{d}_{\theta_k}(n,k) = | \sum_{m=0}^{H-1} \mathbf{h}(m)\mathbf{v}_{\theta_k}(n-m)e^{-j (2 \pi k m/ H)}|^2,
\end {equation}
where $\mathbf{v}_{\theta_k}=\sum_{l=r_l}^{r_u}\mathbf{r}_{\theta_k}(l,q)$ is obtained by collapsing the range dimension beginning  from lower range bin $r_l$ to highest range bin $r_h$. Tapering window $\mathbf{h}$ of length $H$  is applied to reduce the sidelobes. The spectrograms reveal the the different velocities, accelerations and higher order moments which cannot be easily modeled or assumed to follow specific nonstationary structures \cite{10.1117/12.669003, 295203}. We observe  the  motion of two persons performing sign language in a close proximity with each other at different azimuth angles. We aim to correctly  pair the sign to  the corresponding azimuth angle. This is achieved by jointly processing  the spectrograms,  $\mathbf{v}_{\theta_1}(n,k)$ and $\mathbf{v}_{\theta_2}(n,k)$ which are respectively localized at azimuth angles $\theta_1$ and $\theta_2$.    It is clear that the concurrent sign language of multiple persons are hard to  be  distinguished in azimuth by  only using a single antenna. 

\section{Convolutional Neural Network}
\label{Feature extraction and classification}


In this paper, we limit ASL recognition for two subjects.  The   Micro-Doppler images are generated by localizing each subject in angle according to (\ref{c}). Additionally, a Micro-Doppler image is generated without beamforming by considering only one radar receiver. This would help preserve all information in the sense that some part of the image might be lost after applying spatial filtering. (Fig.\ref{Fig.1}) shows the CNN implementation of multi-person ASL recognition.  The three  Micro-Doppler images are passed through separate networks which are combined later at higher-level layers as shown in  (Fig.\ref{Fig.1}) \cite{9455204}.  The initial layers specific to each Micro-Doppler image  attempt to obtain  specific ASL features as captured from a given direction. The shared higher-level layers across three images, attempt to explain  overall system level concepts such as associating the correct sign to a given subject. The image specific layers, comprise of a three layer CNN, in each layer filter sizes of $3*3$ and $9*9$ were concatenated to take advantage of multilevel feature extraction, followed by max pooling layer \cite{8283539}.  The outputs of each  network is flattened and concatenated before it is fed to the shared layers. The shared layers are comprised of fully connected layers followed by dropout layers. \\
The overall data collection, preprocessing   can be described by the following steps. 
\subsection{Data Collection and Preprocessing}
The bandwidth of the 77 GHz sensor is set to 4GHz. The two participants sat on chairs facing a  radar systems with the radial angles of 30$^0$ and -30$^0$. The radar systems were positioned at a distance of 2 meters from the participants. The output transmission power of the RF sensor was 40 mW.
\begin{itemize}
 \item  PRI is set to 1 ms, and each data example is observed over the time period of 4 s, resulting in $Q=4000$ slow time samples.
 \item  ADC sampling rate is 512 ksps, rendering 512 fast time samples per PRI, resultantly the length of data vector is $N=3276800$. 
 \item  The  received data $\mathbf{s}(n,m) \in C^{N*M}$, is collected through $M=4$ element receive array, with an inter-element spacing of $\lambda/2$ ($\lambda$ is the wavelength corresponding to the operating frequency), therefore the dimensionality of received raw data matrix is $3276800\times4$. 
 \item  Beamforming is performed on the raw data matrix, resulting in a spatially filtered $\mathbf{x}({\theta_k})$ vector of dimensions $3276800\times1$.  Two such vectors are generated in the directions of each motion ${\theta_1}$ and ${\theta_2}$.
 \item  Each vector $\mathbf{x}({\theta_k})$
is reshaped into a $512\times12000$ matrix. After applying columnwise DFT,  and identifying the range bins of interest, the corresponding rows are summed together, resulting in  $\mathbf{v}_{\theta_k}=\sum_{l=r_l}^{r_u}\mathbf{r}_{\theta_k}(l,q)$, which is of dimensions $12000\times1$. 
 \item  A combined spectrogram and two spectrograms after beamforming $\mathbf{d}_{\theta_1}$ and $\mathbf{d}_{\theta_2}$, each of dimensions $128\times128$ is obtained, where the window length is 128. 
 \end{itemize}


\begin{itemize}
 
\item The spectrograms  are then passed onto  CNN classifier  for training /prediction. 
 \end{itemize}

\begin{table}[h]
\centering
\caption{Confusion matrix, 3 Combined Signs for 9 Total Classes}
\label{table:1}
\vspace{1mm}
\begin{tabular}{ | m{4.5em} | m{3.2em}| m{3.2em} | m{3.2em} | m{3.2em}| m{3.2em} | m{3.2em} | m{3.2em}| m{3.2em} | m{3.2em} |} 
\hline
Predicted v. Actual & Class-1 (B-B) & Class-2 (B-C)  & Class-3 (B-D) & Class-4 (C-B) & Class-5 (C-C) & Class-6 (C-D) & Class-7 (D-B) & Class-8 (D-C) & Class-9 (D-D)\\ 
\hline
Class-1 (B-B) & 100.0\% & 0\%  & 0\%  & 0\%  & 0\%  & 0\%  & 0\%  & 0\%  & 0\% \\ 
\hline
Class-2 (B-C) & 0\% & 100\% & 0\%  & 0\%  & 0\%  & 0\%  & 0\%  & 0\%  & 0\% \\ 
\hline
Class-3 (B-D) & 0\% & 0\% & 100\%  & 0\%  & 0\%  & 0\%  & 0\%  & 0\%  & 0\% \\ 
\hline
Class-4 (C-B) & 0\% & 0\% & 0\%  & 57.1\%  & 0\%  & 0\%  & 42.9\%  & 0\%  & 0\% \\ 
\hline
Class-5 (C-C) & 0\% & 0\% & 0\%  & 0\%  & 100\%  & 0\%  & 0\%  & 0\%  & 0\% \\ 
\hline
Class-6 (C-D) & 0\% & 0\% & 0\%  & 0\%  & 11.1\%  & 66.7\%  & 0\%  & 0\%  & 22.2\% \\ 
\hline
Class-7 (D-B) & 0\% & 0\% & 0\%  & 33.3\%  & 0\%  & 0\%  & 66.7\%  & 0\%  & 0\% \\ 
\hline
Class-8 (D-C) & 0\% & 0\% & 0\%  & 0\%  & 12.5\%  & 0\%  & 0\%  & 87.5\%  & 0\% \\ 
\hline
Class-9 (D-D) & 0\% & 0\% & 0\%  & 0\%  & 0\%  & 0\%  & 0\%  & 0\%  & 100\% \\ 
\hline
\end{tabular} 
\end{table} 

\hfill \break
\vspace{3mm}


 \section{Experimental Results} \label{Experimental Results}

\begin{figure}[!t]
	\centering
	\includegraphics[height=3in, width=6in]{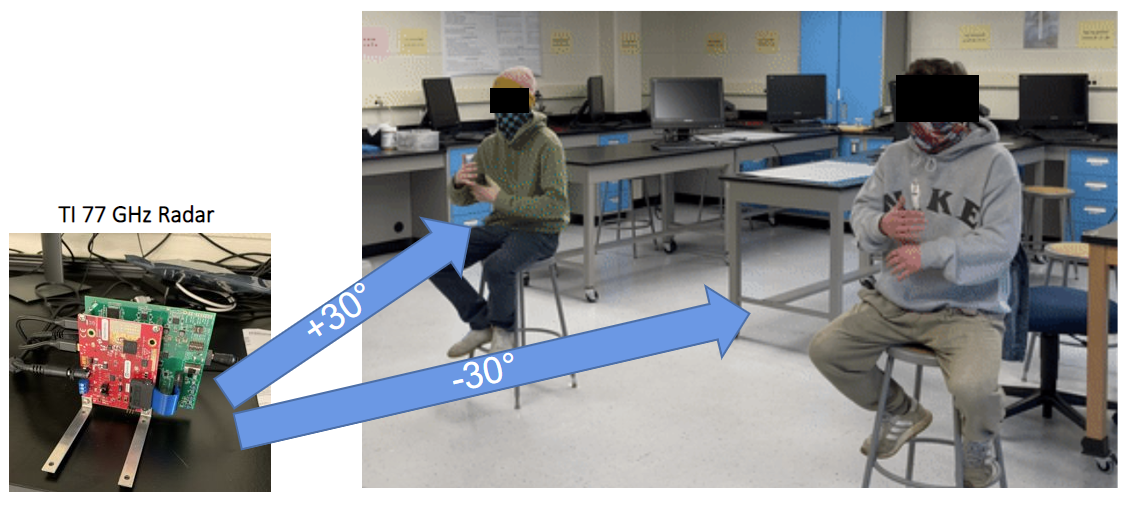}
	\caption{Two Subjects performing ASL signatures simultaneously (Breathe+Breathe)}
	\label{Fig.2}
\end{figure}

\begin{figure}[!t]
	\centering
	\includegraphics[height=2.1in, width=3.5in]{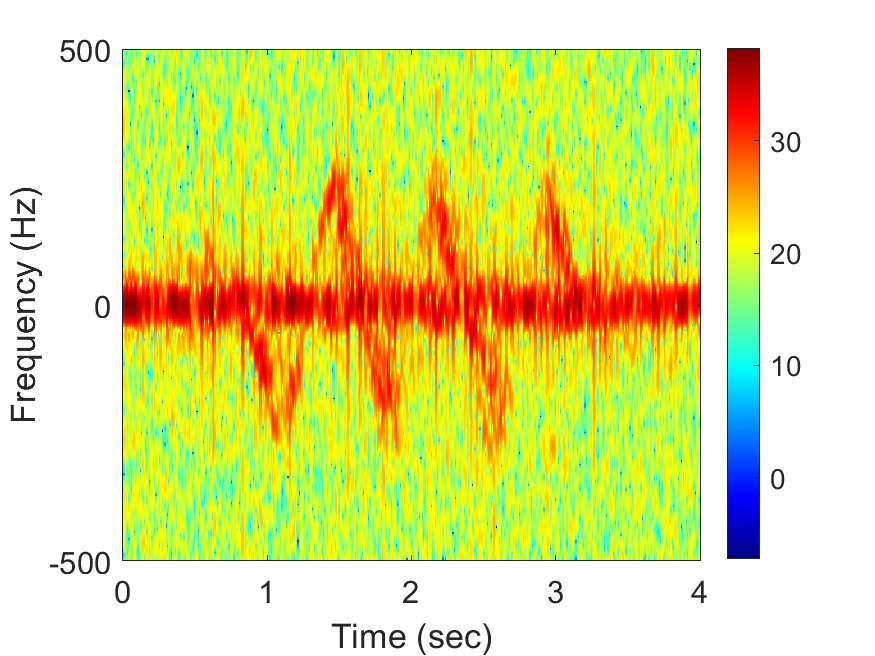}
	\caption{Time Frequency signature of Drink+Breathe signs processed through a single antenna (without beamforming)}
	\label{Fig.3}
\end{figure}

\begin{figure}[!t]
	\centering
	\includegraphics[height=2.3in, width=3.5in]{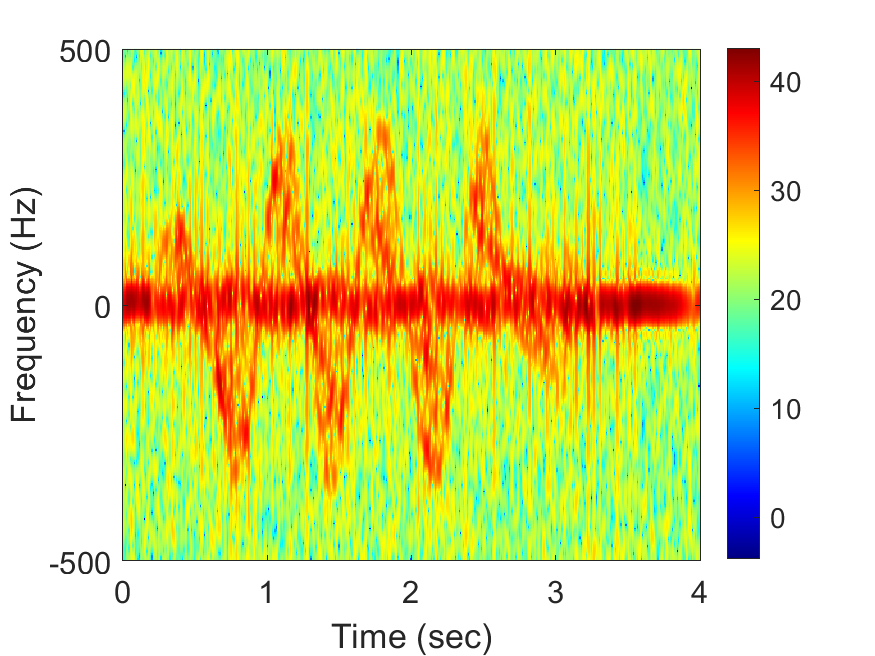}
	\caption{Time Frequency signature of Drink+Breathe signs processed through beamformer pointing towards $\theta_1$ (Showing Breathe).}
	\label{Fig.4}
\end{figure}

\begin{figure}[!t]
	\centering
	\includegraphics[height=2.5in, width=3.5in]{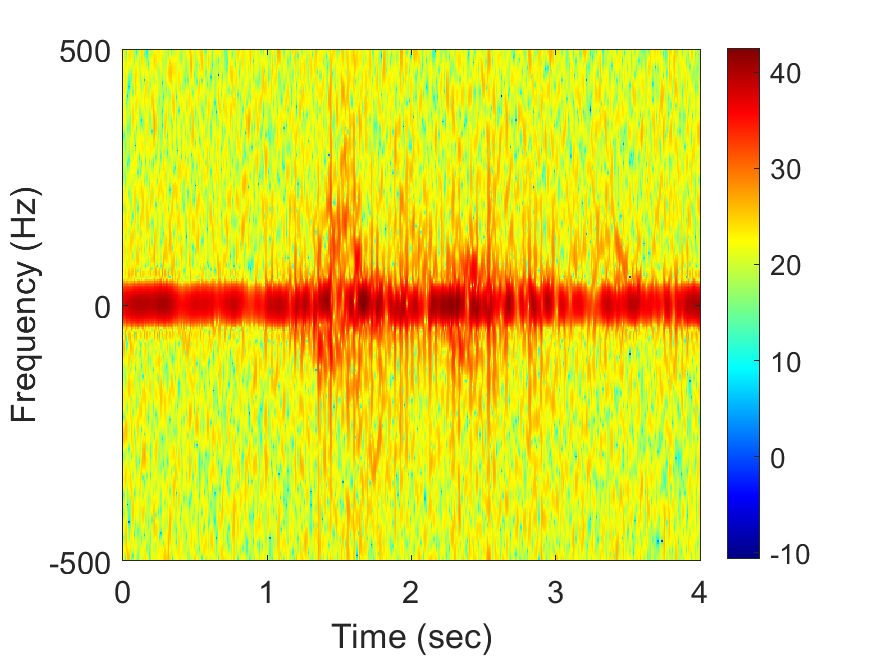}
	\caption{Time Frequency signature of Drink+Breathe signs processed through beamformer pointing towards $\theta_2$ (Showing Drink).}
	\label{Fig.5}
\end{figure}

In this section, we show that the radar system is capable of separating the  Micro-Doppler spectrograms of the two signs such that   convolutional neural network is able to adequately classify them. We consider two persons performing  three different signs, breathe (B), come (C), and drink (D), resulting in nine  classes. Each combination of signs was collected twenty times for two pairs of subjects totaling 360 individual samples. During collection, the radar was centered between two subjects who  sat facing forward at approximately a 30 degree angle from the radar as shown in Fig.\ref{Fig.2}. Radar returns from both subjects was separated during the processing step by taking advantage of delay and sum beamformer. For example, Fig.\ref{Fig.3} shows a combined spectrogram of simultaneously signing drink and breathe signs which are processed through a single antenna (without beamforming). 
Figs. \ref{Fig.4} and \ref{Fig.5} show the time frequency signature of Drink and Breathe signs processed through beamformer pointing towards $\theta_1$ and $\theta_2$ respectively. It is clear that the spectrograms are similar to the individual signs, however, there is some component of the leakage from the other direction.


The radar system used in the experiments is  AWR2243 from Texas Instruments having four receiver and three transmitters.  The center frequency is 77 GHz, whereas the bandwidth is 5 GHz. For the CNN classifications, the training set consists of 80 percent of the samples for each class, and each sample are three 128 × 128 images (one combined image and two beamformed images ). The rest of the data set is used for testing. The combined image is entirely in temporal domain without performing beamforming. The second image is generated by steering the beamformer to $\theta_1$. Likewise, the third image is generated by changing the steering angle to $\theta_2$. Thus, for each motion class three 128× 128 × 20 matrices  are generated. All the three  beamformed TF images are passed separately through a three layer CNN. The three networks are fused  at the fully connected layers at the end. For training, all network weights are optimized in a supervised fashion. 

 The confusion matrix for neural nets sign classification is shown in Table \ref{table:1}, depicting the correspondence between the actual class and the classified class. It shows the neural nets prediction on the horizontal axis and the actual class that corresponded to the Micro-Doppler spectrogram on the vertical axis. This gives a visual representation of which signs are being incorrectly classified as other signs. Each of the classes are labeled with their corresponding sign combination for example BB stands for the sign combination such that both participants are performing  Breathe signature. The values indicate the percentage in which the neural net predicted a class, for example Class-7 (DB) was correctly predicted 66.7\% of the time and was predicted as Class-4 (CB) 33.3\% of  the time. All of these smaller measurements of correct and incorrect classifications provide us with and overall accuracy for the trained neural net. Utilizing only 360 individual samples in the neural net we were able to attain an overall classification accuracy of 88.89\%.




\section{Conclusion}
\label{Conclusion}

In this paper, we introduced an approach that observes the 
time frequency representation of  radar returns from different azimuth angles.  We provided an effective means to discern combinations of multiple ASL signs performed simultaneously at different angular directions from the radar. The proposed approach successfully   maps the ASL signs to the corresponding angular locations and is  found to be effective when the spectrograms  are not completely separable in angle.  

\section{ACKNOWLEDGMENT}
\label{Problem Formulation}

The authors would like to thank Andrew West, Eli Kennedy, and LaJuan Washington Jr. for their assistance with data collection.
\bibliographystyle{IEEEtran}
\bibliography{references}

\end{document}